\documentclass[final,1p,times,twocolumn]{elsarticle}
\usepackage{amssymb}
\usepackage{graphicx,amssymb,amsmath,amsthm,amsfonts,epsfig}
\journal{Nuclear Physics B}
\begin{document}

\begin{frontmatter}



\title{Thermodynamic Product Formula for Ho\v{r}ava Lifshitz Black Hole}


\author{Parthapratim Pradhan}
\ead{pppradhan77@gmail.com}

\address{ Department of Physics, Vivekananda Satavarshiki Mahavidyalaya
(Affiliated to Vidyasagar University), Manikpara, Jhargram, West Midnapur,
West Bengal~721513, India}

\begin{abstract}
We examine the thermodynamic properties of inner and outer horizons in the background of Ho\v{r}ava Lifshitz 
black hole. We compute the \emph{horizon radii product, the surface area product, the entropy product,
the surface temperature product, the Komar energy product and the specific heat product} for both the horizons. 
We show that surface area product, entropy product and irreducible mass product are
\emph{universal}(mass-independent) quantities, whereas the surface temperature product, Komar energy product 
and specific heat product are \emph{not universal} quantities because they all depend 
on mass parameter. We further study the stability of such black hole by computing the specific 
heat for both the horizons. It has been observed  that under certain condition the black hole 
possesses second order phase transition.
\end{abstract}

\end{frontmatter}

\section{Introduction}

There has been a revival of interest in the physics of thermodynamic product formulae in recent years due to the work 
of Ansorg et. al.\cite{ah08,ah09}(see also \cite{mv13,pp14,ac79,ac81}) from the gravity
site and due to the work of Cveti\v{c} et. al. \cite{cgp11} (see also\cite{fl97,cr12,sd12}) from the
string theory site. The main interest therein is the area product formula and the entropy product formula of
event horizon and Cauchy horizons. For example, let us consider a regular axisymmetric and stationary spacetime of
Einstein-Maxwell gravity with surrounding matter then the area product formula of event horizon(${\cal H}^{+}$)
and Cauchy horizons(${\cal H}^{-}$) is\cite{ah09}
\begin{eqnarray}
\frac{{\cal A}_{+} {\cal A}_{-}}{(8\pi)^2} &=& J^2+\frac{Q^4}{4} ~.\label{prKN}
\end{eqnarray}
and consequently the entropy product formula of ${\cal H}^\pm$ are
\begin{eqnarray}
\frac{{\cal S}_{+} {\cal S}_{-}}{(2\pi)^2} &=& J^2+\frac{Q^4}{4} ~.\label{prKN1}
\end{eqnarray}
In the absence of Maxwell gravity, these product formulae reduce to the following form\cite{ac79}:
\begin{eqnarray}
\frac{{\cal A}_{+} {\cal A}_{-}}{(8\pi)^2} &=& J^2 ~.\label{prK}
\end{eqnarray}
and
\begin{eqnarray}
\frac{{\cal S}_{+} {\cal S}_{-}}{(2\pi)^2} &=& J^2 ~.\label{prK1}
\end{eqnarray}
In the above formulae, the common point is that they all are independent of the mass, so-called the
ADM(Arnowitt-Deser-Misner) mass of the background spacetime. Thus they all are universal quantities in
this sense. Now if we incorporate the BPS states, the area product formula should be\cite{fl97,cgp11}
\begin{eqnarray}
\frac{{\cal A}_{+} {\cal A}_{-}}{\left(8\pi {\ell _{pl}}^2\right)^2}  &=& \sqrt{N_{1}}\pm\sqrt{N_{2}}
= N , \,\, N\in {\mathbb{N}}, N_{1}\in {\mathbb{N}}, N_{2} \in {\mathbb{N}} ~.\label{ppl}
\end{eqnarray}
where $\ell _{pl}$ is the Planck length.

It is a well known fact that certain BH possesses inner horizon  or Cauchy horizon(CH) in addition to the
outer horizon or event horizon. Thus there might be a relevance of inner horizon in BH thermodynamics
to understanding the microscopic nature of inner BH entropy in comparison with the outer BH entropy. It is
also true that CH is a blue-shift region whereas event horizon is a red-shift region by its own right.
Furthermore the CH is highly unstable due to the exterior perturbation\cite{sc83}. Despite the above
features, the CH horizon is playing an important role in BH  thermodynamics.

Thus in this note we wish to study various thermodynamic products of Kehagias-Sfetsos(KS) BH\cite{ks09} in
Ho\v{r}ava Lifshitz(HL) gravity\cite{ph9a,ph9b,ph9c}. We have considered both the inner horizon and
outer horizons to  understanding the microscopic nature of BH entropy. We compute various thermodynamic
products like horizon radii product, surface area product, BH entropy product, surface temperature 
product and Komar energy product. By computing the specific heat, we also analyze the stability of 
such BHs.

The plan of the paper is as follows. In Sec. 2, we shall describe the basic properties of the HL BH and 
shall compute various thermodynamic products. In this section there are two subsections. In first subsection, 
we compute irreducible mass product for inner horizon and outer horizons. In second subsection, we discuss the 
stability analysis for this BH by computing the specific heat and also derived the product of 
specific heat for both the horizons. Finally, we conclude our discussions in Sec. 3.

\section{Ho\v{r}ava Lifshitz BH:}
In 2009, Ho\v{r}ava\cite{ph9a,ph9b,ph9c} proposed  a field theory model for a UV complete theory of gravity which is a
non-relativistic renormalizable theory of gravity and reduces to Einstein's general relativity at
large scales for the dynamical coupling constant $\lambda=1$.
Introducing the ADM formalism where the metric can be written as
\begin{eqnarray}
ds^2 &= & -N^2 dt^2 +g_{ij}(dx^{i}-N^{i}dt)(dx^{j}-N^{j}dt)~.\label{adm}
\end{eqnarray}
and for a spacelike hypersurface with a fixed time, its extrinsic curvature $K_{ij}$ is given by
\begin{eqnarray}
K_{ij} &= &\frac{1}{2N}(\dot{g_{ij}}-\nabla_{i}N_{j}-\nabla_{j}N_{i})~.\label{adm1}
\end{eqnarray}
where a dot denotes a derivative with respect to $t$ and covariant derivatives defined with
respect to the spatial metric $g_{ij}$.
The generalized action for Ho\v{r}ava Lifshitz theory is given by
$$
S = \int dt d^3x\sqrt{g}N \big[ \frac{2}{\kappa^2}\big( K_{ij}K^{ij}-\lambda K^{2} \big)+
\frac{\kappa^2 \mu^2(\Lambda_{w} R-3\Lambda^2_{w})}{8(1-3\lambda)}+
\frac{\kappa^2 \mu^2(1-4\lambda)}{32(1-3\lambda)}R^2
$$
\begin{eqnarray}
-\frac{\kappa^2}{2w^4}\big(C_{ij}-\frac{\mu w^2}{2}R_{ij} \big)
\big(C^{ij}-\frac{\mu w^2}{2}R^{ij} \big)+\mu^4 R \big]~.\label{adm2}
\end{eqnarray}
where $\kappa^2, \lambda, \mu, w$ and $\Lambda$ are constant parameters and the cotton
tensor $C_{ij}$ is defined by
\begin{eqnarray}
C^{ij} &=& \epsilon^{ikl} \nabla_{k}\big(R^{j}_{l}-\frac{1}{4}\epsilon^{ikj}\partial_{k}R \big)
~.\label{cotton}
\end{eqnarray}
Comparing the action to that of general relativity, one could find that the speed of light,
Newton's constant and the cosmological constant are
\begin{eqnarray}
c &=& \frac{\kappa^2 \mu}{4}\sqrt{\frac{\Lambda_{w}}{1-3\lambda}} \\
G &=& \frac{\kappa^2}{32 \pi c} \\
\Lambda &=& \frac{3}{2}\Lambda_{w} ~.\label{cg}
\end{eqnarray}
respectively. It may be noted that when $\lambda=1$, the first three terms in Eq. (\ref{adm2})
reduces to the usual one of Einstein's general relativity. It should also be noted that $\lambda$
is a dynamic coupling constant and for $\lambda >\frac{1}{3}$, the cosmological constant must be a
negative one. However, it could be made a positive one if we give a following transformation like:
$\mu\rightarrow i\mu$ and $w^2\rightarrow -iw^2$.

In this work, we emphasized the BH solution in the limit of $\Lambda_{w} \rightarrow 0$. For this
reason, we set $N^{i}=0$ and therefore in order to get the spherically symmetric solution we have
considered the metric ansatz\cite{lmp09,ym09,ks09,mk10,cc9a,cc9b}:
\begin{eqnarray}
ds^2 &= & -N^{2}(r) dt^2 + \frac{dr^2}{g(r)} +r^2(d\theta^2+ \sin^2\theta d\phi^2)~.\label{anstz}
\end{eqnarray}
In order to find the solution, putting the metric ansatz (\ref{anstz}) into the action and we have
the reduced Lagrangian given by
$$
{\cal L} =  \frac{\kappa^2 \mu^2 N}{8(1-3\lambda) \sqrt{g}}\big[(2\lambda -1)\frac{(g-1)^2}{r^2}
-2\lambda \frac{g-1}{r}g'+ \frac{g-1}{2}g'^2
$$

\begin{eqnarray}
 -2\omega(1-g-rg')  \big] ~.\label{as1}
\end{eqnarray}

where $\omega=\frac{8\mu^2(3\lambda-1)}{\kappa^2}$. Since we are interested in this work
to investigate the case $\lambda=1$ i.e. $\omega=\frac{16 \mu^2}{\kappa^2}$. Thus we get the solution
of the metric:
\begin{eqnarray}
N^2(r) &=&g= 1-\sqrt{4 {\cal M} \omega r+\omega^2 r^4} +\omega r^2 .~\label{hl1}
\end{eqnarray}
where ${\cal M}$ is an integration constant related to the mass parameter.
Thus the static, spherically symmetric solution in this case is given by
\begin{eqnarray}
ds^2 &= & -g(r) dt^2 + \frac{dr^2}{g(r)} +r^2(d\theta^2+ \sin^2\theta d\phi^2).~\label{hl}
\end{eqnarray}
For $r\gg (\frac{{\cal M}}{\omega})^{\frac{1}{3}}$, we obtain the usual behavior of a Schwarzschild
BH.

The BH horizons correspond to $g(r)=0$:
\begin{equation} 
r_{\pm}={\cal M} \pm \sqrt{{\cal M}^2 -\frac{1}{2\omega}}.~\label{hl2}
\end{equation}
As is $r_{+}$ corresponds to event horizon and  $r_{-}$ corresponds to Cauchy horizon. Their product
yields
\begin{equation}
r_+ r_- = \frac{1}{2\omega} .~\label{hl3}
\end{equation}
which is clearly independent of mass.

The area of this BH is given by
\begin{eqnarray}
{\cal A}_{\pm} &=& \int^{2\pi}_0\int^\pi_0 \sqrt{g_{\theta\theta} g_{\phi \phi}}d\theta d\phi=4 \pi r_{\pm}^2 \\
                  &=& 4\pi\left(2{\cal M}r_{\pm}-\frac{1}{2\omega} \right)\\
                  &=& 4 \pi \Big[ 2\mathcal{M}^2 -\frac{1}{2\omega} \pm 2\mathcal{M} \sqrt{\mathcal{M}^2-\frac{1}{2\omega}} \Big]
                  ~. \label{M5}
\end{eqnarray}
and their product yields
\begin{eqnarray}
{\cal A}_+{\cal A}_- &=&  \frac{4\pi^2}{\omega^2}.~
\end{eqnarray}
It indicates that the area product formula of KS BH is clearly independent of BH mass. Thus the 
area product formula in HL gravity is universal in nature.

It should be noted that using Eq. (\ref{M5}), the mass could be expressed as in terms of area of both the 
horizons:
\begin{eqnarray}
\mathcal{M}^2 &=& \frac{\mathcal{A}_{\pm}}{16 \pi}+\frac{\pi }{4\omega^2 \mathcal{A}_{\pm}}+\frac{1}{4\omega} ~. 
\label{M16} 
\end{eqnarray}

The BH entropy\cite{bk73} computed at ${\mathcal H}^{\pm}$  corresponds to
\begin{eqnarray}\label{M6}
\mathcal{S}_{\pm} &= &\pi \left(2{\cal M}r_{\pm}-\frac{1}{2\omega} \right).
\end{eqnarray}
Thus the entropy product formula is given by
\begin{eqnarray}
\mathcal{S}_+\mathcal{S}_- &=&  \frac{\pi^2}{4\omega^2}
\end{eqnarray}
It is also independent of mass. Thus the entropy product formula is also universal in HL gravity. For our 
record, we also compute other thermodynamic products.

The surface gravity computed at ${\cal H}^{\pm}$ is given by
\begin{eqnarray}
\kappa_{\pm} = \frac{\omega (r_{\pm}- {\cal M})}{1+\omega r_{\pm}^2}.~ \label{M8}
\end{eqnarray}
and their product is given by
\begin{eqnarray}
\kappa_+\kappa_- &=& \frac{2\omega(1-2{\cal M}^2\omega)}{(1+16{\cal M}^2\omega)}. \label{hl9}
\end{eqnarray}
The Hawking\cite{bcw73} temperature on ${\mathcal H}^{\pm}$ reads off
\begin{eqnarray}
T_{\pm} &=& \frac{\omega (r_{\pm}- {\cal M})}{2\pi(1+\omega r_{\pm}^2)}.~\label{M9}
\end{eqnarray}
and their product yields
\begin{eqnarray} 
T_+ T_- = \frac{\omega(1-2{\cal M}^2\omega)}{2 \pi^2 (1+16{\cal M}^2\omega)} .~\label{M12}
\end{eqnarray}
It may be noted that surface gravity product and surface temperature product both depends
on mass thus they are not universal.
Finally, the Komar\cite{ak59} energy computed at ${\cal H}^{\pm}$ is given by
\begin{eqnarray}
E_{\pm} &=& \frac{(r_{\pm}-{\cal M})(4{\cal M}\omega r_{\pm}-1)}{(4{\cal M}\omega r_{\pm}+1)}
.\label{hl10}
\end{eqnarray}
and  their product reads off
\begin{eqnarray}
E_+E_- &=&  \frac{\frac{1}{2\omega}-{\cal M}^2}{1+16{\cal M}^2 \omega}.~\label{hl13}
\end{eqnarray}
which indicates that the product does depend on the mass parameter and thus it is not an universal
quantity.

\subsection{ Irreducible mass product for KS BH in HL gravity:}
Firstly, Christodoulou\cite{cd70} had shown that the irreducible mass
${\cal M}_{\text{irr}}$ of a non-spinning BH which is related to the surface area ${\cal A}$
given by
\begin{eqnarray}
{\cal M}_{\text{irr}, +}^{2} &=& \frac{{\cal A}_{+}}{16\pi}=\frac{{\cal S}_{+}}{4\pi}
~. \label{irrm}
\end{eqnarray}
where `$+$' sign indicates for ${\cal H}^{+}$
and it is now established that this relation is valid for CH also. That means
\begin{eqnarray}
{\cal M}_{\text{irr}, -}^{2} &=& \frac{{\cal A}_{-}}{16\pi}=\frac{{\cal S}_{-}}{4\pi}
~. \label{irrm1}
\end{eqnarray}
Here, `$-$' indicates for ${\cal H}^{-}$.

Equivalently, the expressions for area on ${\cal H}^{\pm}$, in terms of ${\cal M}_{\text{irr}, \pm}$ are:
\begin{eqnarray}
{\cal A_{\pm}} &=& 16 \pi ({\cal M}_{\text{irr}, \pm})^2 ~. \label{M20}
\end{eqnarray}
Thus for KS BH, the product of the irreducible mass at the horizons ${\cal H}^{\pm}$ are:
\begin{eqnarray}
{\cal M}_{\text{irr}, +} {\cal M}_{\text{irr},-}
&=& \frac{1}{8\omega}.~\label{M22}
\end{eqnarray}
The above product is an universal quantity because it does not depend upon the mass
parameter. It should be noted that Eq. (\ref{M16}) can be written as in terms of irreducible mass:
\begin{eqnarray}
{\cal M} &=& {\cal M}_{\text{irr}, \pm}+ \frac{1}{8 \omega {\cal M}_{\text{irr}, \pm}} .~\label{hl23}
\end{eqnarray}

\subsection{Heat Capacity $C_{\pm}$ on ${\cal H}^{\pm}$:}

One can define the specific heat on ${\cal H}^{\pm}$ is given by
\begin{eqnarray}
C_{\pm} &=& \frac{\partial{\cal M}}{\partial T_{\pm}} .~\label{c1}
\end{eqnarray}
which is an important measure to study the thermodynamic properties of a BH. To determine it we first calculate
the partial derivatives of mass ${\cal M}$ and temperature $T_{\pm}$ with respect to $r_{\pm}$ are:
\begin{eqnarray}
\frac{\partial {\cal M}}{\partial r_{\pm}} &=& \frac{2 \omega r_{\pm}^2-1}{4 \omega r_{\pm}^2}.~\label{c2}
\end{eqnarray}
and
\begin{eqnarray}
\frac{\partial T}{\partial r_{\pm}} &=& \frac{\big(1+5\omega r_{\pm}^2-2\omega^2 r_{\pm}^4 \big)}
{8 \pi r_{\pm}^2\big(1+\omega r_{\pm}^2\big)^2} .~\label{c3}
\end{eqnarray}
where the temperature of the BH could be found via the following relation:
\begin{eqnarray}
T_{\pm} &=& \frac{1}{4\pi}\frac{dg(r)}{dr}\mid_{r=r_{\pm}}.
\end{eqnarray}
which comes out
\begin{eqnarray}
T_{\pm} &=& \frac{2\omega r_{\pm}^2-1}{8 \pi r_{\pm}(1+\omega r_{\pm}^2)} .~\label{c4}
\end{eqnarray}
The temperature $T_{\pm}$ of the black hole is positive, zero or negative, depending on whether
$2\omega r_{\pm}^2-1>0$, $2\omega r_{\pm}^2-1=0$ or $2\omega r_{\pm}^2-1<0$ respectively.

Therefore the final expression for heat capacity
$C_{\pm}=\frac{\partial {\cal M}}{\partial r_{\pm}} \frac{\partial r_{\pm}}{\partial T_{\pm}}$
at the ${\cal H}^{\pm}$ becomes:
\begin{eqnarray}
C_{\pm} &=& \frac{2 \pi  }{\omega} \frac{(2\omega r_{\pm}^2-1)\big(1+\omega r_{\pm}^2\big)^2}
{1+5\omega r_{\pm}^2-2\omega^2 r_{\pm}^4}  .~\label{c5}
\end{eqnarray}
Let us analyze the above expression of specific heat for a different regime in the parameter space.

Case I: The specific  heat $C_{\pm}$ is positive when $\omega r_{\pm}^2 >\frac{1}{2}$, in this
case the BH is thermodynamically stable.

Case II: The  specific  heat $C_{\pm}$ is negative when $\omega r_{\pm}^2 < \frac{1}{2}$, in this
case the BH is thermodynamically unstable.

Case III:  The  specific  heat $C_{\pm}$ blows up when $1+5\omega r_{\pm}^2-2 \omega^2 r_{\pm}^4=0 $ i.e.
$r_{\pm}=\pm \sqrt{\frac{5\pm \sqrt{33}}{4\omega}}$, in this
case the BH undergoes a second order phase transition.

Interestingly, the product of specific heat on ${\cal H}^{\pm}$ becomes
\begin{eqnarray}
C_{+} C_{-} &=& \frac{\pi^2}{2\omega^2}\frac{\big(1-2{\cal M}^2\omega \big)\big(1+16{\cal M}^2
\omega \big)^2}{2+13{\cal M}^2\omega-16{\cal M}^4\omega^2}  .~\label{c6}
\end{eqnarray}
It indicates that the product depends on mass parameter and coupling constant $\omega$. Thus
the product of specific heat of both the horizons are not an universal quantity .

It should be noted that in the extremal limit ${\cal M}^2=\frac{1}{2\omega}$, the temperature, Komar energy
and specific heat of the BH becomes zero.

\section{Discussion:}
In this short note, we have studied the thermodynamic features of KS BH in HL gravity. We computed
various thermodynamic product formula for this BH. We observed that the surface area product, BH 
entropy product and irreducible mass product are universal quantities, whereas the surface 
temperature product, Komar energy product and specific heat product are not universal quantities 
because they all are depends on mass parameter. 

Finally, we studied the stability of such black hole by computing the
specific heat for both the horizons. It has been shown that under certain condition the black hole
possesses  second order phase transition.  In summary, these product formulae may somehow help us
a little bit to understanding the microscopic nature of BH entropy both interior and exterior
which is the main aim in quantum gravity.

\bibliographystyle{model1-num-names}

\begin{thebibliography}{99}
\bibitem{ah08} M. Ansorg and J. Hennig, \textit{Class. Quant. Grav.} {\bf 25} {222001} (2008).

\bibitem{ah09} M. Ansorg and J. Hennig, \textit{Phys. Rev. Lett.} {\bf 102} {221102} (2009).

\bibitem{mv13} M. Visser, \textit{Phys. Rev.} {\bf D 88} {044014} (2013).

\bibitem{pp14} P. Pradhan, \textit{The European Physical Journal C} {74} {2887} (2014).

\bibitem{ac79} A. Curir, \textit{ Nuovo Cimento} {\bf 51B } {262} (1979).

\bibitem{ac81} A. Curir, \textit{Gen. Rel. Grav.} {\bf 13 } {12} (1981).

\bibitem{cgp11} M. Cveti\v{c}, G. W. Gibbons and C. N. Pope, \textit{Phys. Rev. Lett.} {\bf 106} {121301} (2011).

\bibitem{fl97} F. Larsen,  \textit{Phys. Rev.} {\bf D 56} {1005} (1997).

\bibitem{cr12} A. Castro and M. J. Rodriguez, \textit{ Phys. Rev.} {\bf D 86} {024008} (2012).

\bibitem{sd12} S. Detournay, \textit{Phys. Rev. Lett. } {\bf 109}, {031101} (2012).

\bibitem{sc83} S. Chandrashekar, \textit{The Mathematical Theory of Black Holes}, Clarendon Press, Oxford (1983).

\bibitem{ks09} A. Kehagias and K. Sfetsos, \textit{Phys. Lett. } {B 678} {123} (2009).

\bibitem{ph9a} P. Horava, \textit{Phys. Rev. Lett. } {\bf 102}, {161301} (2009).

\bibitem{ph9b} P. Horava, \textit{Phys. Rev. } {\bf D 79}, {084008} (2009).

\bibitem{ph9c} P. Horava, \textit{J. High Energy Phys.} {\bf 0903} {020} (2009).

\bibitem{lmp09} H. Lu, J. Mei and C. N. Pope, \textit{Phys. Rev. Lett. } {\bf 103}, {091301} (2009).

\bibitem{ym09} Y. S. Myung, \textit{Phys. Lett. } {B 678} {127} (2009).

\bibitem{mk10} Y. S. Myung and Y. W. Kim, \textit{The European Physical Journal C} {68} {265} (2010).

\bibitem{cc9a} R. G. Cai, L. M. Cao and N. Ohta, \textit{Phys. Rev. D.} {\bf 80} {024003} (2009).

\bibitem{cc9b} R. G. Cai, L. M. Cao and N. Ohta, \textit{Phys. Lett.} {\bf B 679} {504} (2009).

\bibitem{bk73} J. D. Bekenstein, \textit{ Phys. Rev.} {\bf D 7} {2333} (1973).

\bibitem{ak59} A. Komar, \textit {Phys. Rev. } {\bf 113}, 934 (1959).

\bibitem{bcw73} J. M. Bardeen, B. Carter, S. W. Hawking, {\it Commun. Math. Phys.} {\bf 31}, {161} (1973).

\bibitem{cd70} D. Christodoulou, \textit{Phys. Rev. Lett.} {\bf 25} {1596} (1970).
\end{thebibliography}

\end{document}